# Accurate grain boundary plane distributions for textured microstructures from stereological analysis of orthogonal two-dimensional electron backscatter diffraction orientation maps

Martin Folwarczny[a,b,*], Ao Li[a], Rushvi Shah[a], Aaron Chote[a], Alexandra C. Austin[a,b], Yimin Zhu[a,b], Gregory S. Rohrer[c], Michael A. Jackson[d], Souhardh Kotakadi[a,b], Katharina Marquardt[a,b,e]

[a] *Department of Materials, Imperial College London, London, SW7 2AZ, UK*
[b] *Department of Materials, University of Oxford, Oxford, OX1 3PH, UK*
[c] *Department of Materials Science and Engineering, Carnegie Mellon University, Pittsburgh, PA 15213, USA*
[d] *BlueQuartz Software, 400 S Pioneer Blvd, Springboro, OH 45066, USA*
[e] *The Zero Institute, Holywell House, Osney Mead, Oxford, OX2 0ES, UK*

*\* Corresponding author*
*Department of Materials*
*University of Oxford*
*Parks Road*
*Oxford OX1 3PH*
*United Kingdom*
*Email: martin.folwarczny@materials.ox.ac.uk*

## Abstract

We present a method for obtaining qualitatively accurate grain boundary plane distributions (GBPD) for textured microstructures using a stereological calculation applied to two-dimensional electron backscatter diffraction (EBSD) orientation maps. Stereology, applied to 2D EBSD orientation maps, is currently the fastest method of obtaining GBPDs. Existing stereological methods are not directly applicable to textured microstructures because of the biased viewing perspectives for different grain boundary types supplied from a single planar orientation map. The method presented in this work successfully removes part of this bias by combining data from three orthogonal EBSD orientation maps for stereology. This is shown here to produce qualitatively correct GBPDs for heavily textured synthetic microstructures with hexagonal and tetragonal crystal symmetries. Synthetic microstructures were generated to compare the stereological GBPD to a known ground truth, as the true GBPD could be obtained from a triangular mesh of the full grain boundary network in 3D. The triangle mesh data contained all five macroscopic parameters to fully describe the grain boundary structure. It was observed that our stereological method overestimated the GBPD anisotropy. However, qualitative analysis of the GBPD remains useful. Furthermore, it was found that combining data from three orthogonal sections gives reliable results when sectioning the texture's primary axes.

## 1. Introduction

In the past decade, the study of grain boundary (GB) populations and their properties has gone through a revival [1–5]. It is well-established that grain boundaries govern material properties, which is evident in the constitutional laws that exhibit grain size sensitivity, for example, in yield strength and creep [6], as well as in conductivity and optical properties [7]. Many studies focus on the average grain size as the main parameter related to macroscopic properties, sometimes also including texture analysis. Yet the past two decades have shown that properties of the grain boundaries themselves, like their structure and composition, can be equally influential on the macroscopic properties of the material [3,8,9].

The crystallography of the grain boundary plane fundamentally controls the properties of the grain boundary. Two grain boundaries displaying the same misorientation between adjacent grains and nominally displaying the same Σ (inverse lattice coincidence) value can connect adjacent crystals on very different crystal planes, leading to different properties, such as different grain boundary migration rates [10] or space charge potentials [9].

Furthermore, the GB plane fundamentally links GB energy to various other GB properties, including oxide ion conductivity [11,12]. A study by Han et al. [13] highlights that GBs of a specific misorientation occur in multiple crystallographic variations with vastly different energies, termed GB states, even when considering misorientations about only one axis and limited to symmetric grain boundaries. It is important to note that special low energy grain boundaries have the fewest number of available GB states, and therefore, the non-special grain boundaries exhibit a greater variation in their GB state and energy.

Grain boundary plane distributions (GBPD) are a measure for the distribution of the crystallographic orientation of grain boundaries with respect to their respective host grain. One may consider the GBPD as the average crystal shape in a polycrystalline material. The GBPD is commonly displayed on a stereographic projection in units of multiples of random (or uniform) distribution (MRD), showing the relative areas of different crystallographic planes on the grain boundary [14].

By studying the whole distribution of all grain boundary types present in the material, the bias from studying only a few boundaries, which might be special in terms of properties, structure or chemistry, can be removed. This is particularly important, seeing that non-special grain boundaries show more variable grain boundary states and respond strongly to material processing [13]. Furthermore, non-special GBs have been found to be more frequent than special Σ boundaries in materials such as alumina and zirconia [15], likely impacting material properties more significantly than the few special GBs.

Grain boundary character, as defined using five macroscopic degrees of freedom from misorientation $\Delta g(\varphi_1, \Phi, \varphi_2)$ and plane normal $n(\theta, \phi)$ [16], cannot be obtained from two-dimensional maps measured by electron backscatter diffraction (EBSD) directly [17]. Although the boundary plane normal orientation is not fully known, it can be computed statistically using stereology, for example, using code written at Carnegie Mellon University in the Rohrer group [17,18].

This stereological analysis works well for samples with no preferred crystal orientations, i.e., non-textured materials. This is because the analysis relies on supplied observations of individual types of grain boundaries from random observation perspectives, as it is a statistical method. Taking two-dimensional sections of very strongly textured materials supplies biased observation perspectives for each grain boundary type. Furthermore, two-dimensional sections of textured microstructures contain a biased subset of grain boundary types. This causes nonuniform sampling and thus further biases computed stereological GBPDs of textured materials.

Textured materials constitute a large proportion of modern materials studied; consequently, extending the stereological GBPD analysis approach is required. Being able to obtain accurate grain boundary character measurements using two-dimensional EBSD orientation maps would be a great simplification compared to three-dimensional orientation map techniques, such as synchrotron X-ray diffraction [19], serial sectioning [20–22] or transmission electron microscopy [23,24]. These techniques can directly access all five grain boundary character parameters, but are too expensive, inaccessible, and difficult for routine applications.

One potential approach to limit the texture bias in stereology is to take multiple differently oriented sections, as proposed by Saylor [17]. While this must lead to a more representative subsection of boundaries and observation perspectives supplied to the stereological method, the number and orientation of sections required to obtain an accurate representation of the grain boundary character are difficult to predict.

Here, we benchmark methods for computing the GBPD from two-dimensional data by sampling simulated 3D microstructures for which the true GBPD is known. The GBPD was calculated from plane sections of simulated textured microstructures with hexagonal and tetragonal symmetries by combining data from three orthogonal plane sections. The results were then compared to the GBPD calculated from the entire 3D microstructure, representing the true GBPD.

## 2. Method

### 2.1. Synthetic microstructure generation

We combined microstructure simulation of a textured material using DREAM.3D [25,26] with an analysis of these microstructures as if they were an experimental sample mapped by EBSD. To benchmark the stereological method, two case studies with different crystal systems were used, both based on real-world textured materials.

The first dataset simulated was Zircaloy-4 in the α-Zr phase with hexagonal crystal symmetry, often used as a cladding material for fuel rods in nuclear pressurised water reactors (PWR) [27,28]. A typical rolling texture for zirconium alloys was applied, with {10-10} aligned with the rolling direction, {11-20} aligned with the transverse direction, and {0001} aligned with the normal direction [29].

The second simulated dataset was a tetragonal phase of zirconia ($ZrO_2$), which is one of the phases that develop in the corrosion resistant oxide layer on Zircaloy-4 in PWRs, likely stabilised by the large pressure in a PWR [30,31]. The texture simulated was a typical (001) fibre texture [32].

The intensities of textures applied to the microstructures simulated in this project were deliberately much higher than what would be expected in real materials. For typical rolling textures, the maximum intensities of orientation distribution plots are less than 10 in multiples of random distribution (MRD) [29,32]. The microstructures simulated for this project had maximum intensities reaching 50 MRD. The strong texture made the resulting GBPDs more predictable. When simulating a microstructure in DREAM.3D, first the grain shapes were simulated to fill the space of the synthetic volume, and only afterwards were the crystal orientations assigned to the already formed grains to satisfy the orientation distribution function and the disorientation distribution function [25]. Therefore, it was not guaranteed that the crystal orientation would exactly match the orientation of the grain elongation for each grain. Using extremely strong textures reduced the potential mismatch between the orientations of the crystal and the grain shape for each grain, making the results more predictable.

The simulation steps, called filters in DREAM.3D, used to generate the synthetic microstructures, are shown in Table 1. Note that in this microstructure simulation, GBs are not attributed with grain boundary energy – they simply follow the shape we define during the simulation set up. In other words, they will, on average, depict elliptical grain shapes.

*Table 1: DREAM.3D filters used for generating synthetic microstructures.*

| Filter number | Filter name |
|---|---|
| 1 | StatsGenerator |
| 2 | Initialize Synthetic Volume |
| 3 | Establish Shape Types |
| 4 | Pack Primary Phases |
| 5 | Find Feature Neighbors |
| 6 | Match Crystallography |

Using DREAM.3D, we generated the textured synthetic mono-phased microstructures. First, we used filter 1 to set the phases, including the crystal symmetry. The grain size was specified using a lognormal distribution by setting the mean, standard deviation, and maximum and minimum grain sizes specified by the number of standard deviations allowed in each direction. The parameters used are shown in Table 2.

In filter 1, the texture was also defined. Grain shape was defined by ellipsoid axis ratios (A:B:C). The orientation of the elongated grains with respect to the sample reference frame was specified using Euler angles in the Bunge convention. The crystallographic texture was set by specifying orientations via Euler angles in the Bunge convention and assigning each with a weight and the width of the distribution around the absolute value. The parameters are shown in Table 2.

*Table 2. DREAM.3D microstructure simulation parameters used for each dataset. Grain size distributions were set in DREAM.3D using lognormal distributions, by specifying μ, σ and number of standard deviations in both directions from the mean. The real grain size is defined as the equivalent sphere diameter (ESD). Grain size distribution μ and σ parameters are defined as unitless in DREAM.3D, and the user is instructed to adjust them until the estimated grain ESD reaches the desired value. The grain size metric used in MTEX was equivalent circle diameter (ECD).*

| | Setting | Hexagonal | Tetragonal | Tetragonal: Texture strength test | Tetragonal: Misaligned sections test |
|---|---|---|---|---|---|
| | **DREAM.3D input** | | | | |
| Filter 1 | Estimated mean equivalent sphere diameter (ESD) (μm) | 14.90 | 10.23 | 10.23 | 10.23 |
| Filter 1 | Lognormal grain size dist. μ | 2.55 | 2.2 | 2.2 | 2.2 |
| Filter 1 | Lognormal grain size dist. σ | 0.55 | 0.5 | 0.5 | 0.5 |
| Filter 1 | No. standard deviations towards min. grain size | 4.0 | 2.0 | 2.0 | 2.0 |
| Filter 1 | No. standard deviations towards max. grain size | 2.5 | 1.5 | 1.5 | 1.5 |
| Filter 1 | A:B:C ellipsoid axis ratio | 3:1:1 | 4:1:1 | Trial 1 - 3:2.5:2<br>Trial 2 - 4:3:1<br>Trial 3 - 6:3:1<br>Trial 4 - 8:3:1 | 3:2:1 |
| Filter 1 | Ellipsoid orientation distribution function (Euler Angles, Bunge) | $\phi_1 = 0°$<br>$\Phi = 90°$<br>$\phi_2 = 0°$ | $\phi_1 = 0°$<br>$\Phi = 90°$<br>$\phi_2 = 90°$ | $\phi_1 = 0°$<br>$\Phi = 90°$<br>$\phi_2 = 90°$ | $\phi_1 = 17°$<br>$\Phi = 90°$<br>$\phi_2 = 42°$ |
| Filter 1 | Crystallographic orientation distribution function (Euler Angles, Bunge) | $\phi_1 = 20°, 30°, 40°$<br>$\Phi = 0°$<br>$\phi_2 = 0°$ | $\phi_1 = 0°, 10°, 20°, 30°, 40°, 50°, 60°, 70°, 80°$<br>$\Phi = 0°$<br>$\phi_2 = 0°$ | $\phi_1 = 0°, 10°, 20°, 30°, 40°, 50°, 60°, 70°, 80°$<br>$\Phi = 0°$<br>$\phi_2 = 0°$ | $\phi_1 = 107°$<br>$\Phi = 48°$<br>$\phi_2 = 0°, 10°, 20°, 30°, 40°, 50°, 60°, 70°, 80°$ |
| Filter 2 | No. voxels in 3D | 700x700x700 | 501x501x501 | 501x501x501 | 501x501x501 |
| Filter 2 | Step size (μm) | 1.4 | 1.0 | 1.0 | 1.0 |
| | **Simulation results (MTEX analysis)** | | | | |
| | Grain size (ECD) mean (μm) | 15.41 | 8.91 | Trial 1 - 9.55<br>Trial 2 - 9.31<br>Trial 3 - 8.99<br>Trial 4 - 8.72 | 9.24 |
| | Grain size (ECD) median (μm) | 12.00 | 7.16 | Trial 1 - 7.61<br>Trial 2 - 7.46<br>Trial 3 - 7.21<br>Trial 4 - 6.95 | 7.54 |
| | Grain size (ECD) mode (μm) | 7.27 | 4.62 | Trial 1 - 4.83<br>Trial 2 - 4.80<br>Trial 3 - 4.63<br>Trial 4 - 4.43 | 5.03 |

The sample volume was defined in filter 2 by setting the number of voxels in each direction and the spacing between them. The spacing also represented the step size for EBSD mapping. Orientation maps were exported from the three different sample sections (XY, XZ and YZ) as shown in Figure 1. The Euler angle reference frames were rotated depending on which section was being exported to align the simulated reference frame with the experimental EBSD reference frame. This means that the work protocol and methodology developed here are applicable to EBSD maps obtained from experimental measurements. As shown in Figure 1, the XY section was exported without rotation, a 90° rotation about the X sample reference frame axis was applied in DREAM.3D to the Euler reference frame when exporting the XZ section, and the YZ section required a further rotation of 90° about the new Y sample reference frame axis.

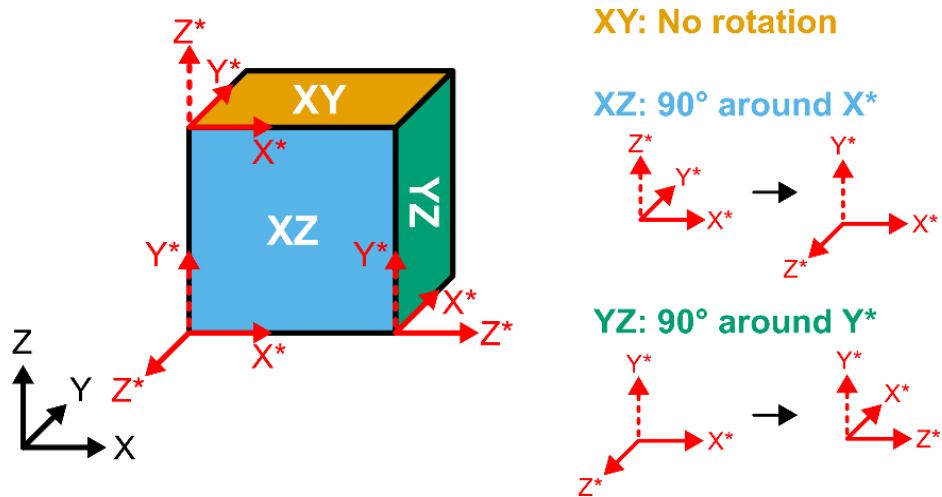

*Figure 1. Schematic of the rotations performed in the DREAM.3D pipeline to export 2D sections that correspond to sections of an experimental textured sample characterised using EBSD. The black axes define the sample reference system. The green, blue and orange sides of the sample cube are named XY, XZ and YZ, using the two axes defining the respective section in the sample reference frame. To replicate experimentally measured EBSD data, the reference frame was rotated for XZ and YZ sections to keep the coordinate system consistent. IPF maps plotted in OIM Analysis and MTEX were always plotted with respect to the Z direction in the sample reference frame. The directions corresponding to the Z sample direction for each individual section are shown as a dashed red arrow.*

To ensure sufficient statistics for stereology while not creating overly large and time-consuming simulations, multiple sections were exported in each direction, as shown in Figure 2. The microstructures simulated using DREAM.3D are spatially uniform, so using multiple sections is statistically equivalent to sampling larger sections, if the spacing between the sections is chosen to be larger than the maximum grain size to avoid sampling artefacts. The maximum ellipse long axis length, determined using MTEX, was used as the measure for setting the section spacing.

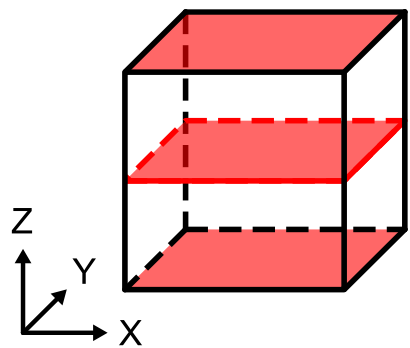

*Figure 2. Visualisation of taking multiple slices in the XY direction. The same is done for the other directions.*

Subsequently, the planar sections of the simulated microstructure were exported from DREAM.3D. The exported files contained a list of Euler angles, each corresponding to a different voxel in the DREAM.3D spatial coordinate system. A MATLAB script was written to convert these files into orientation maps in the .ang format. This is the standard format used in TSL OIM Analysis by EDAX.

Step size and sample dimensions set in DREAM.3D were used to assign each Euler angle to a physical coordinate in space. A typical header was then added to the new .ang files with information about the specific phase simulated.

### 2.2. Texture validation

The first step in the analysis of the synthetic orientation maps was performed using the MTEX toolbox for MATLAB [33]. To verify the input parameters for the DREAM.3D simulations, the crystallographic texture, grain size distribution, grain shape and orientations were extracted and compared to the DREAM.3D input. Additionally, the results from MTEX were compared with the display in the ParaView software [34] to ensure consistency in reference frame conventions used. This is a suitable comparison as ParaView uses the same reference frame conventions as DREAM.3D.

Standard functions provided by MTEX were used to plot inverse pole figure (IPF) maps and orientation distribution function (ODF) pole figures to analyse the crystallographic texture of the simulated microstructure. Figure 3 displays three orthogonal sections plotted with IPF colour with respect to the sample Z direction. The comparison of the input ODF plots from DREAM.3D and the pole figures plotted in MTEX from the exported sections can be found in Figure 4.

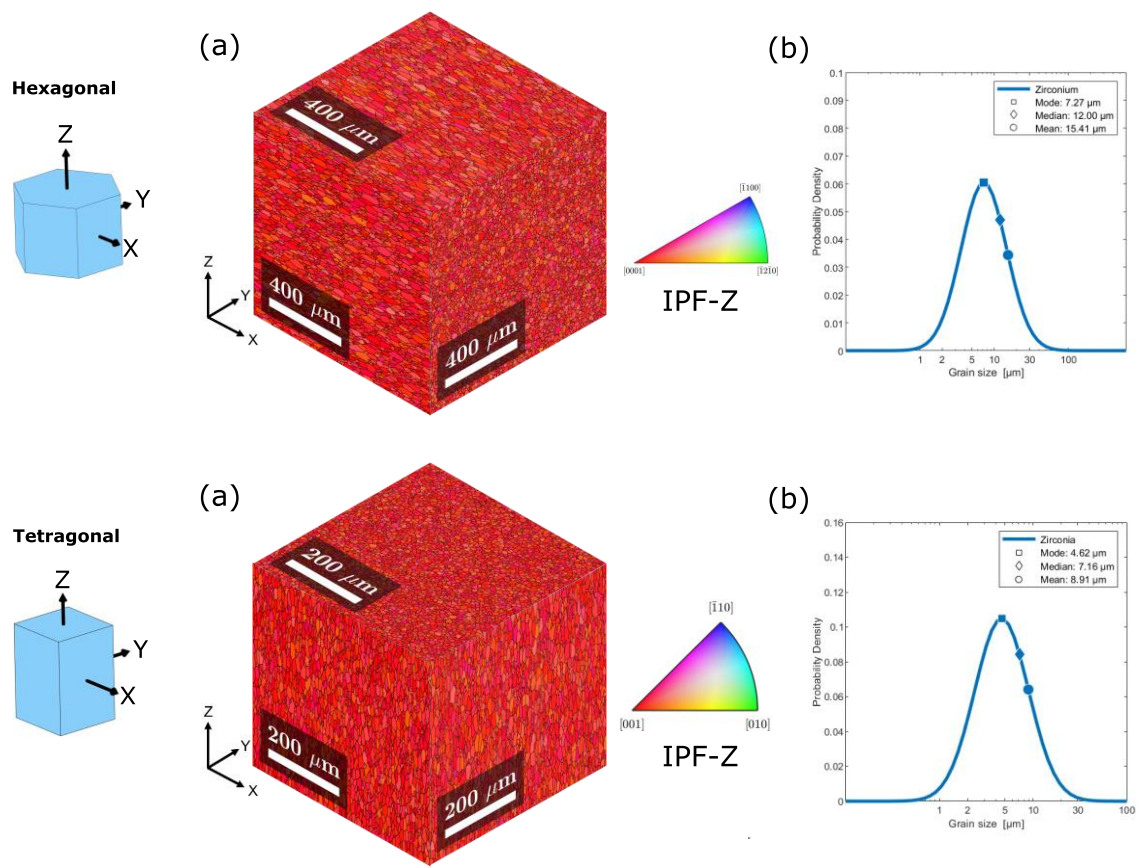

*Figure 3. Confirming texture and the spatial and rotational consistency of microstructures simulated using DREAM.3D. (a) Combined IPF maps for XY, YZ and XZ sections plotted in MTEX, all orientations along the Z sample reference frame axis. (b) Grain size analysis, including mode, median and mean (values listed in Table 2).*

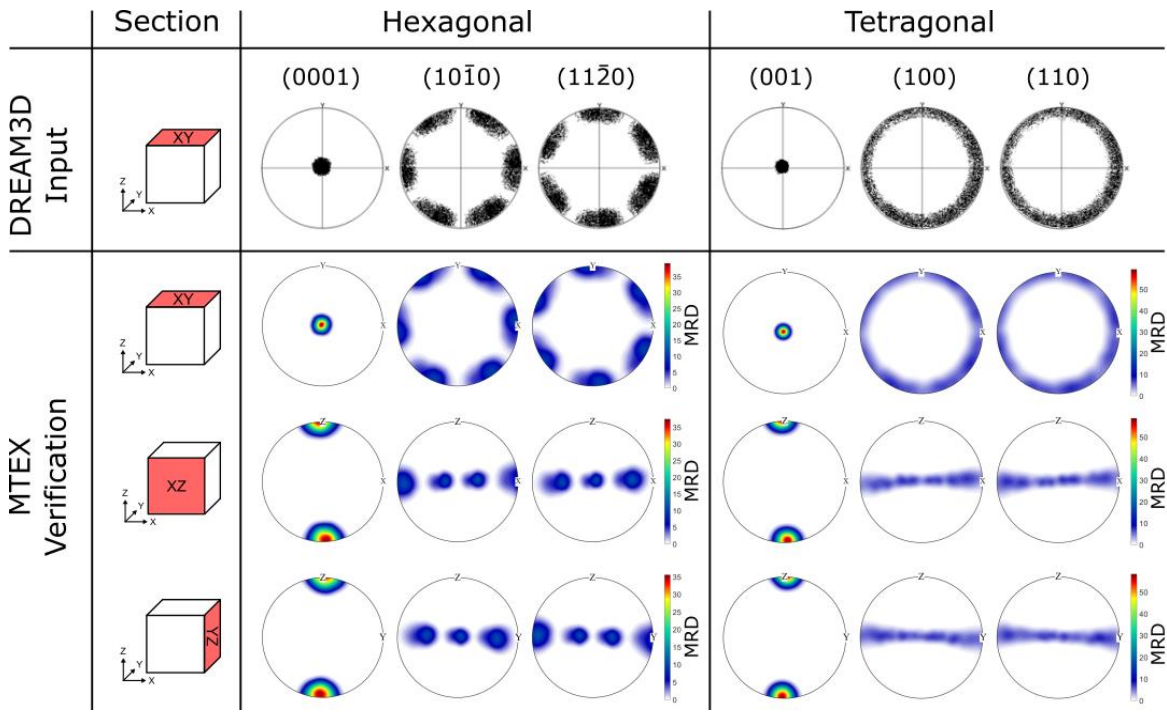

*Figure 4. Confirming the texture of the simulated hexagonal and tetragonal microstructures using ODF pole figures. Both the DREAM.3D input and the MTEX analysis of the simulated microstructure are shown. MTEX plots are shown for all three orthogonal sections (XY, XZ and YZ) individually. The plots are in good agreement.*

The grain size, shape and orientation properties of the microstructure were also analysed using MTEX, fitting grains with ellipses. Grain size was defined using the equivalent circle radius function in MTEX. Grain diameters were calculated and fitted with a lognormal distribution. The resulting grain size distributions are shown in Figure 3.

The grain shape was analysed in two ways: aspect ratio and characteristic shape function. Results are plotted in Figure 5 for the hexagonal microstructure and Figure 6 for the tetragonal. The aspect ratios were calculated by comparing the lengths of the ellipse long and short axes. The characteristic shape function is provided by MTEX and sums over grain boundary segment orientations. Ellipse long axis orientation is plotted on a polar histogram.

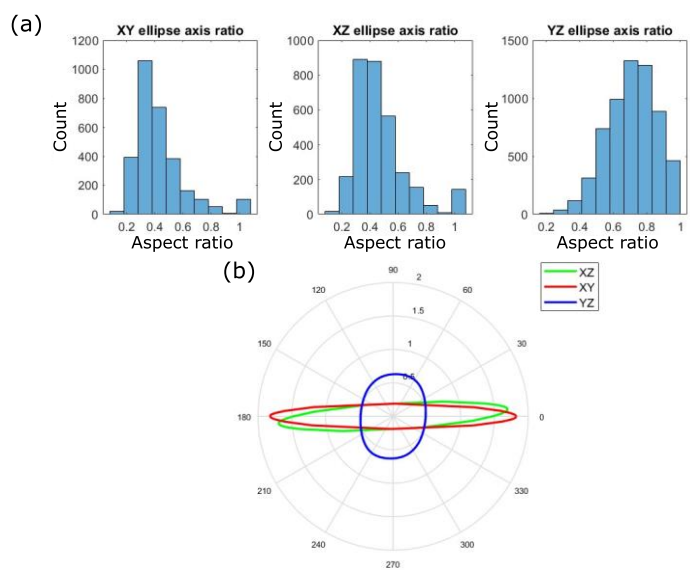

*Figure 5. Grain shape analysis for hexagonal microstructure plotted by MTEX for the three orthogonal sections individually. (a) Ellipse short to long axis ratios describing the level of elongation of grains. (b) The characteristic shape of grains from all three sections separately, normalised for area.*

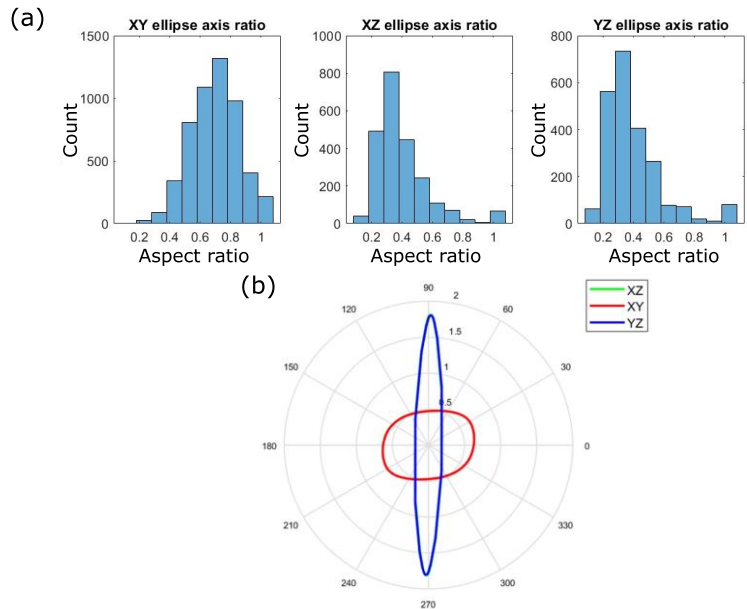

*Figure 6. Grain shape analysis for tetragonal microstructure plotted by MTEX for the three orthogonal sections individually. (a) Ellipse short to long axis ratios describing the level of elongation of grains. (b) The characteristic shape of grains from three sections separately, normalised for area. XZ and YZ characteristic shape functions overlap nearly perfectly, resulting in the XZ data being poorly visible.*

## 2.3. Additional datasets

The effect of a less symmetrical grain shape was also tested. This was achieved by simulating microstructures with different lengths of the ellipsoid short axes (i.e. 4:3:1 compared to 4:1:1). Furthermore, microstructures with varying levels of grain elongation were simulated to test the dependence of the bias in the stereological method on the magnitude of the applied texture. These simulations were performed using the same settings as the original tetragonal symmetry dataset, with only the ellipsoid axes ratios varied. The exact parameters used can be found in Table 2 in the "Tetragonal: Texture strength test" column. The same texture verification method was used for these datasets as for the original hexagonal and tetragonal datasets.

Lastly, the effect of alignment of the sections taken through the sample with respect to the texture was investigated. To this end, a misaligned version of the same microstructure for the original dataset with tetragonal crystal symmetry was simulated. Both the ellipsoid grain orientations and the crystallographic orientations were tilted away from the sample reference frame in the same way. In the previous datasets, the sample reference frame, ellipsoid grain axes and crystallographic texture were all perfectly aligned with each other. The parameters used are shown in Table 2 in the "Tetragonal: Misaligned sections test" column, and the texture analysis is in Figure 7.

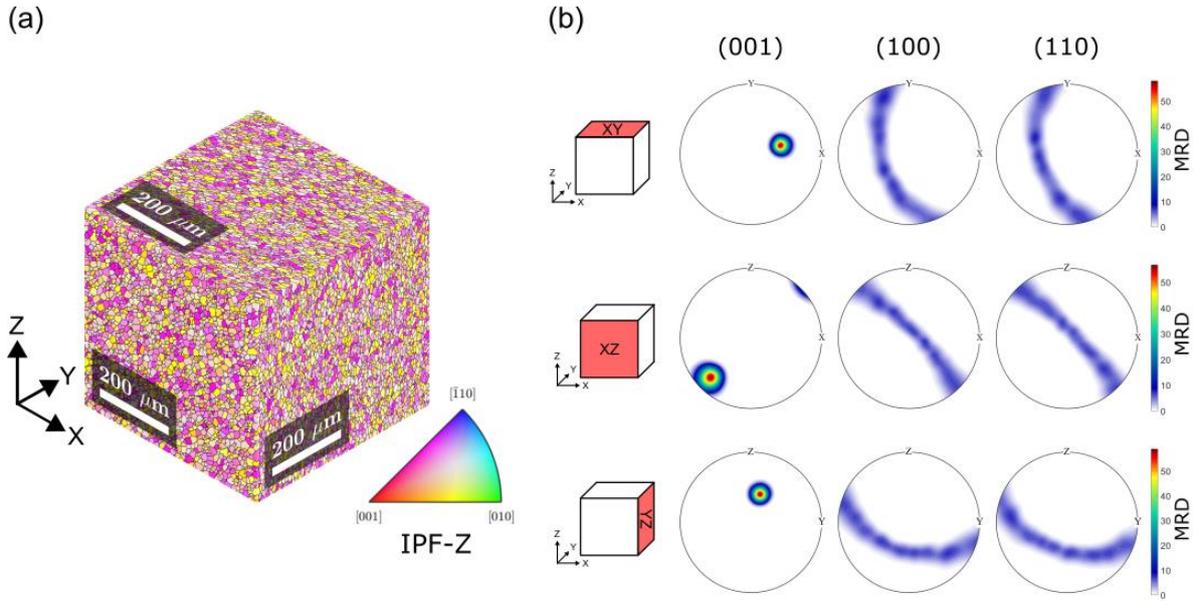

*Figure 7: Tetragonal crystal structure dataset with the texture misaligned with respect to the sample directions X, Y, Z. This simulates taking the orthogonal sections from a material at random, without aligning them with the texture. A 3:2:1 ellipsoid axis ratios were used to simulate a grain shape with uniaxial symmetry. (a) IPF maps of the three sections with respect to the Z sample direction. (b) ODF plots of the three different sections, displaying the misaligned texture with respect to sections made (in contrast to the aligned tetragonal texture in Figure 4).*

## 2.4 Plotting grain boundary plane distributions

The resulting grain boundary plane distributions (GBPDs) were plotted using code developed in the Rohrer group [14,17,18]. This code uses grain boundary segments calculated from grain boundary traces as input. To extract the grain boundary segments, our orientation maps (.ang files created in MATLAB) were imported into TSL OIM Analysis 8 software. When manually creating the orientation map files (.ang), coordinates were set up with the origin of the coordinate system in the lower left corner, x-direction increasing to the right, y-direction increasing to the top, and z-direction pointing out of the page. Thus, a rotation of 90° around the A3 axis was applied to all sections after importing into TSL OIM Analysis 8, following guidance on converting between coordinate systems [35]. ODF pole figures were plotted in OIM Analysis and compared to those from DREAM.3D and MTEX to verify the importing procedure. TSL OIM Analysis 8 was solely used to extract grain boundary segments without any additional data cleaning or processing, as the synthetic microstructures in this project were free from artefacts normally encountered in experimental EBSD data. From the grain boundary segments, GBPDs were calculated for the three orthogonal sections separately as well as combined (Figure 8 and Figure 9).

To compute the true GBPD for the simulated microstructure, the voxelated boundaries between the grains in the simulation had to be replaced by a smooth mesh of triangles. The creation of the triangular meshes was carried out in DREAM.3D using the filters found in Table 3. The resulting ground truth GBPDs were calculated using the 3D data analysis function of the code developed by the Rohrer group [17,18,36].

*Table 3: DREAM.3D filters used for converting simulated grain boundaries into a mesh of triangles.*

| Filter number | Filter name |
|---|---|
| 1 | Quick Surface Mesh |
| 2 | Laplacian Smoothing |
| 3 | Generate Triangle Normals |
| 4 | Generate Triangle Areas |
| 5 | Export GBCD Triangles File |

# 3. Results

Figure 8 shows the GBPDs produced for the dataset with hexagonal crystal symmetry, simulated using parameters from the "Hexagonal" column in Table 2. The stereologically calculated GBPDs from single 2D sections and all three 2D sections combined are compared to the ground truth GBPD calculated from the 3D grain boundary structure of the simulated microstructure.

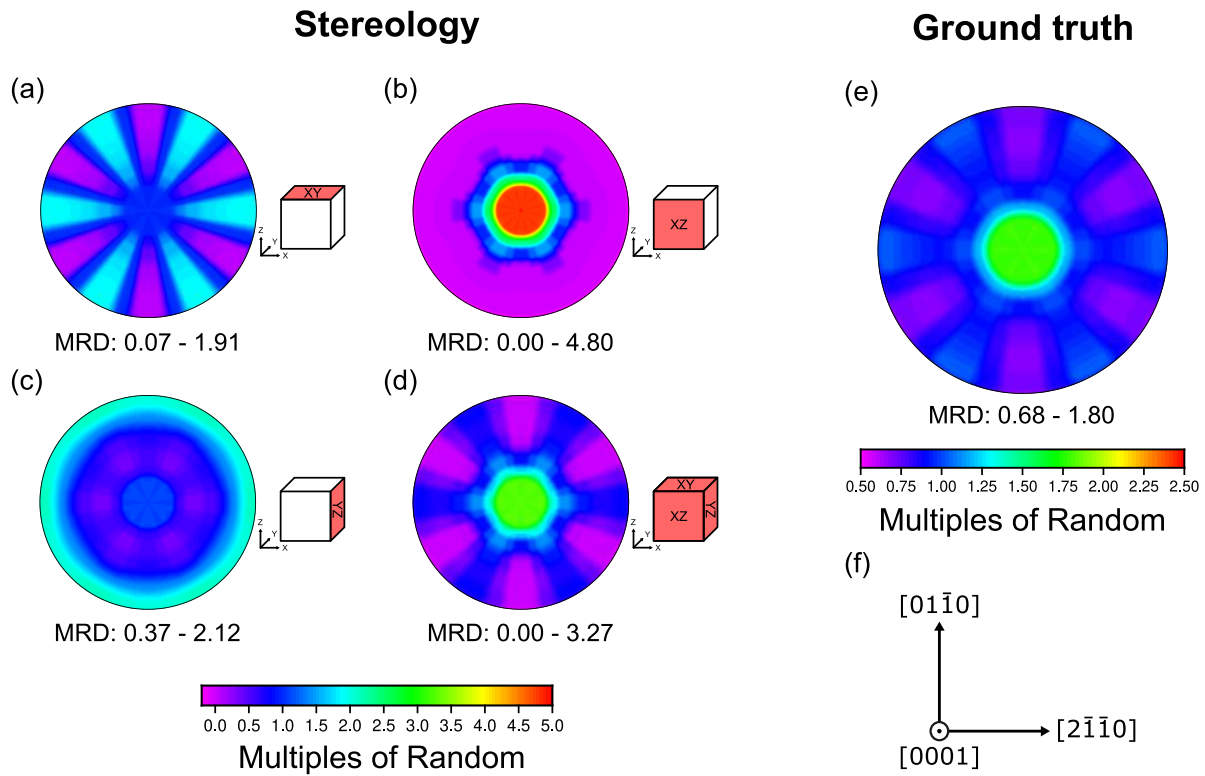

*Figure 8. GBPD plots for the simulated microstructure with hexagonal crystal symmetry. Stereological GBPDs from the individual sections (a)-(c), and all three sections combined (d). All four plots are plotted using the same colour scale. (e) Ground truth GBPD obtained from triangle meshing the 3D grain boundary network, plotted with a different colour scale than the stereological GBPDs. (f) Axes used for plotting the stereographic projections in (a)-(e).*

The GBPD for the XY section (Figure 8 a) shows preference for planes in the $2^{nd}$ order prismatic zone, mainly {11-20}. The $1^{st}$ order prismatic zone planes, particularly {10-10}, are not expressed. The XZ section (Figure 8 b) displays a strong preference for the {0001} basal planes, reaching 4.8 MRD. The GBPD for the YZ section (Figure 8 c) preferentially displays planes parallel to [0001] direction, described as {hki0} planes where "h", "k" and "i" are integers. The $1^{st}$ order {10-10} and the $2^{nd}$ order {11-20} prismatic planes and higher index planes are represented.

The combined GBPD for the three orthogonal sections (Figure 8 d) displays a preference for the {0001} basal planes, reaching 3.27 MRD. $1^{st}$ order prismatic zone planes appear very little in the entire dataset, while $2^{nd}$ order prismatic zone planes are relatively randomly distributed. It represents the average of the GBPD of the XY, XZ and YZ sections. The GBPD describes an average crystal habitus, in our case, an elongated ellipsoid of biaxial symmetry, longest in the X direction, or in other words, the shape of a rice grain.

The ground truth GBPD obtained from triangle meshing the 3D grain boundary structure is shown in Figure 8 e. Similar to the stereological GBPD from all three sections combined, it displays a clear preference for the {0001} basal planes reaching 1.8 MRD, a below average representation of the $1^{st}$ order prismatic planes and an above average representation of $2^{nd}$ order prismatic planes. The level of anisotropy in the GBPD, characterised by the difference between maximum and minimum intensity of the distribution, is much lower for the ground truth result than the stereology results.

Figure 9 shows the resulting GBPDs for the dataset with tetragonal crystal symmetry, simulated with parameters in the "Tetragonal" column in Table 2. Both the stereologically calculated and the ground truth GBPDs are shown.

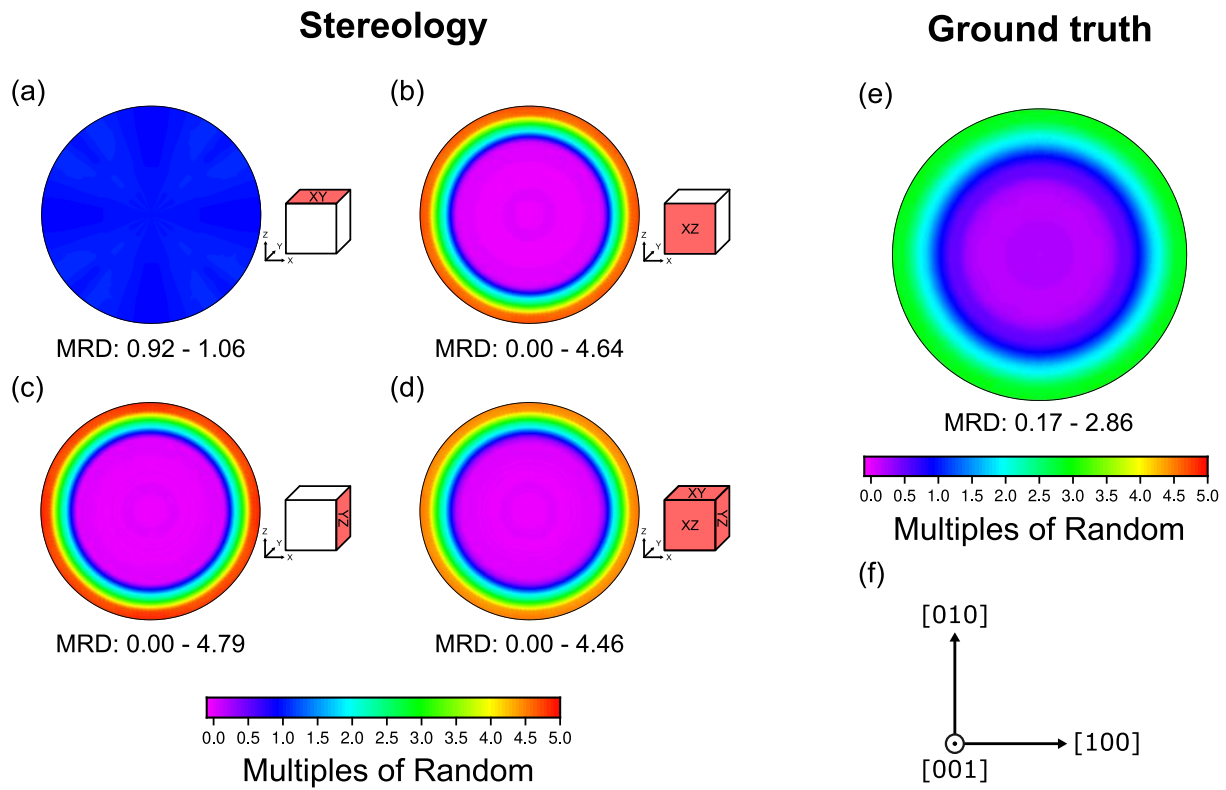

*Figure 9. GBPD plots for the simulated microstructure with tetragonal crystal symmetry. Stereological GBPDs from the individual sections (a)-(c), as well as all three sections combined (d). All four plots are plotted using the same colour scale. (e) Ground truth GBPD obtained from triangle meshing the 3D grain boundary network, also with the same colour scale. (f) Axes used for plotting the stereographic projections in (a)-(e).*

The GBPD for the XY section (Figure 9 a) shows close to random distribution of GB planes. Sections XZ (Figure 9 b) and YZ (Figure 9 c) have a very similar GBPD, showing a strong preference for planes parallel to [001] direction, here referred to as {hk0} where "h" and "k" can be any arbitrary integer. These planes are around 4.5 times more common compared to a random distribution. As the planes deviate from being parallel to [001] direction, they become increasingly rare. The purple section of the GBPD in the middle signifies planes appearing scarcely in the dataset, namely the {001} family.

The combined GBPD from all three orthogonal sections in Figure 9 (d) is an averaged plot of the individual section GBPDs. The pattern shows the same features as XZ and YZ because of their high intensity compared to XY, but with a lower maximum intensity. This is most clearly visible on the red outside circle, which is slightly less intense than in individual plots for XZ and YZ.

Figure 9 e shows the ground truth GBPD obtained from triangle meshing the grain boundaries. It also displays a preference for the {hk0} planes, with intensity reaching 2.86 MRD, and diminishing frequency of planes when deviating from {hk0}. The level of anisotropy in the ground truth GBPD is lower than in the stereological GBPD from combined sections.

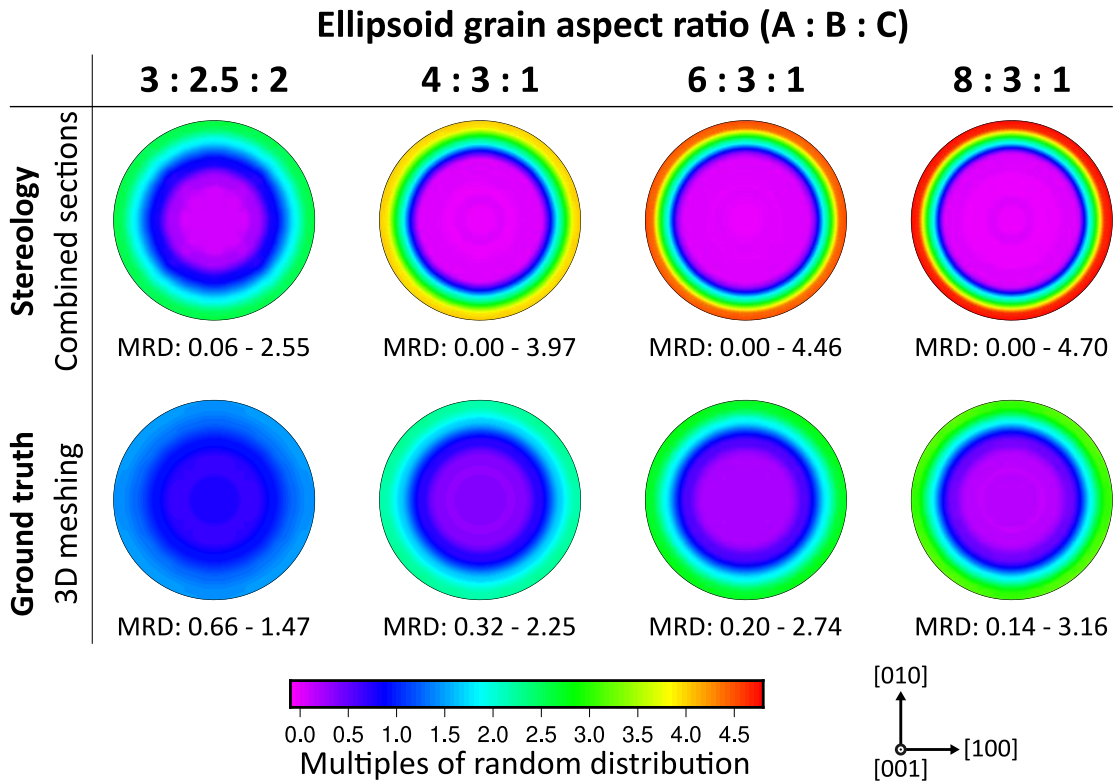

*Figure 10. GBPD results for tetragonal crystal symmetry microstructures with changing the ellipsoid grain shape, as described by the aspect ratio. The top row shows GBPDs calculated stereologically using combined data from three orthogonal sections, and the bottom row shows the ground truth GBPDs calculated from triangular meshes of the grain boundaries. All plots use the same colour bar shown. The minimum and maximum MRD values are included for all GBPDs shown. Both the stereological and ground truth GBPDs increase in maximum intensity with increasing relative length of the ellipsoid A axis.*

The stereologically calculated GBPDs for microstructures with varying ellipsoid long axis (A) length are shown in Figure 10. These datasets correspond to parameters shown in the column "Tetragonal: Texture strength test" in Table 2. Both the stereological GBPDs calculated from combined data from three orthogonal sections and the ground truth GBPDs obtained using triangular meshing are shown for comparison. The patterns obtained show similar features to the original tetragonal dataset GBPDs shown in Figure 9, displaying preference for {hk0} planes and reduced frequency of {001} planes.

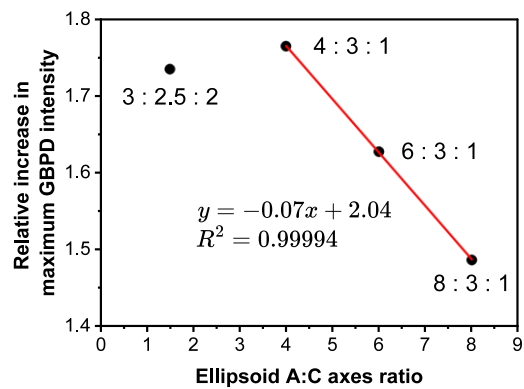

*Figure 11: Relative increase in the maximum intensity of the stereological GBPDs compared to the ground truth GBPDs for tetragonal crystal symmetry microstructures with changing grain elongation, characterised here by the ellipsoid aspect ratio A:C. The values used here are taken from the GBPDs shown in Figure 10. A linear relationship is observed when only the A axis is varied, but breaks when the B axis is also changed. Datapoints are labelled with the full A:B:C aspect ratio.*

It can be seen in Figure 10 that the maximum intensity and level of anisotropy of both the stereologically computed and the ground truth GBPDs become greater with increasing level of elongation (or aspect ratio) of the grains, i.e. with increasing strength of the texture. The data presented in Figure 11 shows that the ratio of maximum intensities of stereological compared to ground truth GPBDs followed a linear trend with the ellipsoid grain aspect ratio for datasets where only the A:C ratio was varied and B:C was kept constant. The one dataset where B:C was varied did not follow the linear trend.

Figure 12 shows the results from applying the stereological method of combining data from three orthogonal sections in a scenario where the sections measured are misaligned from the ellipsoid grain axes and the crystallographic texture. The simulation parameters for this microstructure can be found in column "Tetragonal: Misaligned sections test" in Table 2 and the resulting microstructure is shown in Figure 7.

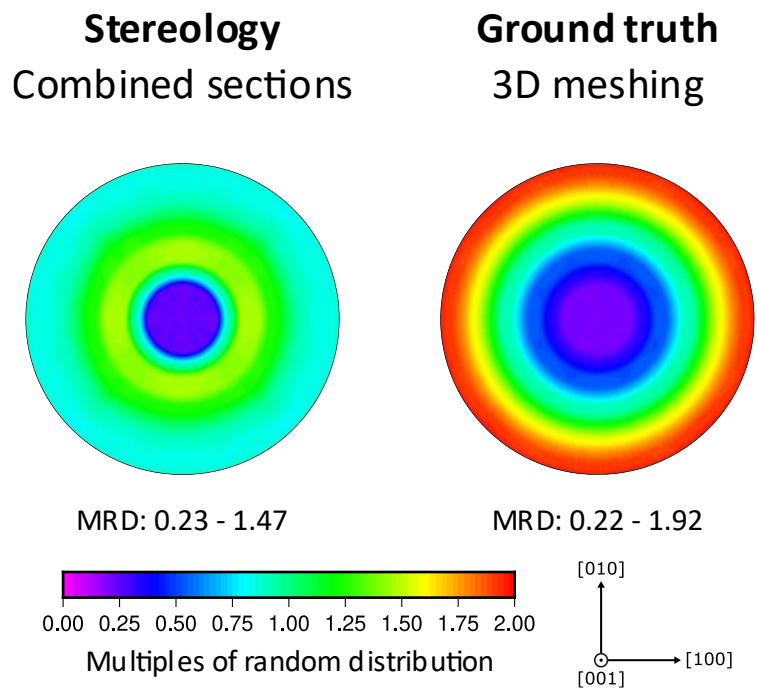

*Figure 12: Comparison of the stereologically calculated GBPD to ground truth for the tetragonal dataset with sections misaligned from the texture primary axes, as shown in Figure 7. The ground truth GBPD was obtained by triangle meshing the simulated microstructure, while the stereological GBPD was calculated using combined segments from three orthogonal sections.*

The ground truth GBPD shows a similar pattern to the previous tetragonal crystal symmetry microstructures, with the sections aligned with the texture primary axes, where {hk0} planes dominate the GBPD. The stereological GBPD from combined sections displays a different distribution, where the most frequent grain boundary plane normals are tilted away from the [001] direction around 48°. The level of anisotropy in the stereological GBPD is also lower than in the ground truth GBPD, counter to what was observed when the texture axes and observation directions were aligned.

## 4. Discussion

Before evaluating the obtained GBPDs, note that the simulated microstructures are non-physical. DREAM.3D does not consider interfacial energy, so GBPD plots are analysed geometrically only. The resulting patterns originate from the combination of crystallographic texture with grain morphology. This is visualised in Figure 13 for both the hexagonal and tetragonal datasets. The poles of an ellipsoid make up a smaller fraction of the surface area than the equator (the circular sections in Figure 13). Therefore, the crystallographic planes that are oriented in the direction of the poles will make up a smaller fraction of the planes on the grain boundaries than planes oriented towards the equator.

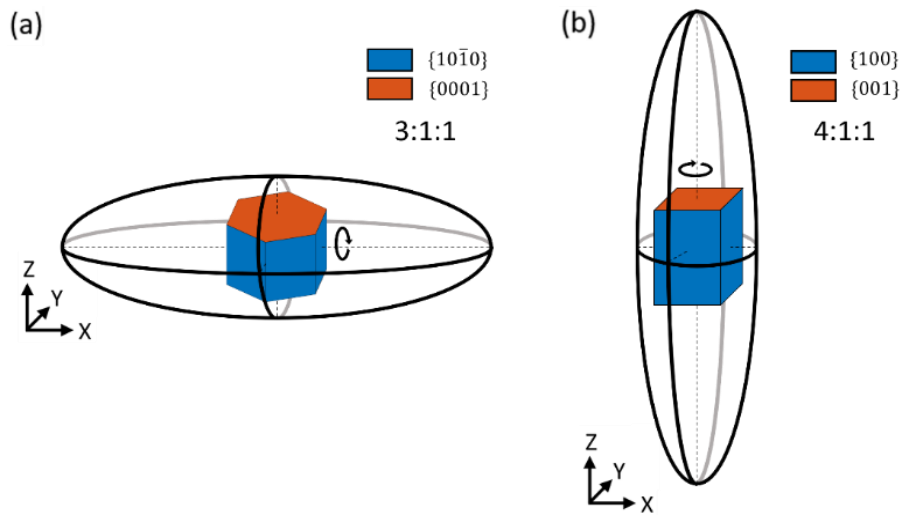

*Figure 13. Schematic visualisation of the combination of crystallographic texture, visualised using a crystallographic unit cell, and ellipsoid grain shape and orientation. Used for justifying the GBPD patterns observed. (a) is a schematic for the hexagonal dataset, (b) is for the tetragonal dataset.*

The combined GBPD for the hexagonal symmetry dataset shown in Figure 8 indicates that basal {0001} planes dominate in the sample. Furthermore, planes in the 2$^{nd}$ order prismatic zone are more common than planes in the 1$^{st}$ order prismatic zone. These results were expected, considering the combined effect of the ellipsoidal grain shape, the ellipsoid orientations, and the crystallographic texture, as shown in Figure 13 a.

The 1$^{st}$ order prismatic planes are aligned with the poles of the ellipsoids as shown in Figure 13 a. Hence, the 1$^{st}$ order prismatic planes make up a smaller fraction of the grain boundaries than planes aligned with the ellipsoid surface at the equator. In the GBPD from only the XY section, the 2$^{nd}$ order prismatic {11-20} planes are aligned with the equator, and in the GBPD from the XZ section only, the basal {0001} planes are aligned with the equator. Therefore, these planes dominate in their respective GBPD plots. The circular pattern with preference for {hki0} planes in the YZ section can be explained by the fact that the ellipsoid cross-sections in this direction are approximately circular, as shown in Figure 5 b. Furthermore, it can be observed that the anisotropy in the YZ sections GBPD is much lower than the other sections, caused by the ellipsoids being sectioned at random heights in the X direction, therefore sampling random boundaries.

In 3D, some of the {10-10} planes are aligned with the ellipsoid poles, while all {0001} planes and some {11-20} planes are aligned with the equator in Figure 13 a. The combined GBPD plot shown in Figure 8 d corresponds well to the simulated crystallographic texture and grain morphology, showing {0001} planes as most common and {11-20} planes more common than {10-10}.

The GBPD calculated stereologically using data from all three sections (Figure 8 d) resembles qualitatively the ground truth GBPD for the hexagonal dataset (Figure 8 e). The stereological method using three orthogonal sections successfully identified the general pattern, showing which GB planes appear with increased or decreased frequency in the microstructure than randomly distributed.

However, the maximum intensity of the ground truth GBPD is lower than that of the stereological one, and the minimum intensity is higher. The stereological method's bias toward overestimating the GBPD's anisotropy requires further evaluation.

For the tetragonal symmetry dataset, the combined GBPD for all sections in Figure 9 shows a strong preference for {hk0} direction. Following the same reasoning as for the hexagonal dataset, these results were expected.

Figure 13 b shows that the poles of the ellipsoids are aligned with {001} planes and the equator with {hk0} planes. The random distribution pattern observed for the XY section is caused by the sectioning of ellipsoids at random heights, producing random MRD. The equal representation of all {hk0} planes in the GBPDs for XZ, YZ and combined data from all three planes stem from the fact that the crystallographic texture only sets {001} planes to be aligned with Z direction, whereas the rotation of the grains around the Z axis is random, producing the circular ODF plot for {100} and {110} planes shown in Figure 4. Overall, the GBPDs calculated stereologically for the individual sections and all three sections combined correspond well to what is expected.

Similar to the results for the hexagonal dataset, it can be seen that the ground truth GBPD presented in Figure 9 (e) shows a pattern similar to the stereological GBPD, combining data from three sections, albeit with different intensities. This shows that for the tetragonal dataset, our stereological method was also able to replicate the general pattern well; again, the anisotropy of the GBPD was overestimated.

In Figure 10, the GBPDs for the datasets testing the effect of texture strength show that with increasing level of elongation of the grains, the intensity of both the stereological and ground truth GBPD increases. This behaviour is expected, as increasing the elongation of the ellipsoid means that the fraction of the surface area associated with the equator of the ellipsoid increases, while the fraction of the surface area associated with the poles remains relatively constant.

An interesting result was that the ratio of maximum intensities of the stereological GBPD compared to the ground truth follows a linear trend when only the A:C ellipsoid axes ratios were varied but breaks for the microstructure where the B:C ratio was also changed, possibly opening a way to use a texture-related factor for a correction algorithm. As mentioned previously, the investigation performed here is based purely on the geometry of the grains, as the microstructure simulation method does not consider grain boundary energy. Therefore, it is not clear whether the same behaviour would be observed in real materials.

It was also shown that the orientation of the three orthogonal sections with respect to the sample's texture is important when applying our stereological method to textured microstructures. The results obtained for the dataset where the texture and the 2D sections were misaligned with respect to each other show that the method of combining data from three orthogonal sections failed to correctly identify the most common GB planes, as demonstrated in Figure 12. Notably, we observe a maximum in the combined GBPD at 48° away from the [001] direction, which coincides with the 48° of tilt applied to the texture in the simulation.

In summary, calculating the GBPDs from three orthogonal sections using stereology alleviates the strong bias in results from a single section of a textured sample. As Saylor outlined, ideally, one would use a large number of randomly oriented sections through a sample to provide a truly random sample of grain boundary planes present [17]. However, this approach would be difficult to implement experimentally, highlighting the practicability of three orthogonal sections until more advanced texture correction becomes available.

# 5. Conclusion

1. It has been shown that plotting GBPDs for textured materials from single section EBSD data often produces highly inaccurate results. Some plots from single sections even failed to recognise the most and least frequently represented grain boundary planes, making the stereological GBPDs from single sections misleading. This result highlights the need to develop alternative methodologies for textured samples.

2. Combining data from EBSD maps of three orthogonal sections alleviates some bias in stereological GBPD measurement of heavily textured samples. The combined GBPD plots from three orthogonal sections showed patterns that corresponded well to the ground truth, albeit with a variation in the MRD values. We suggest that GBPD plots obtained via three orthogonal sections are suitable for qualitative analysis of dominant grain boundaries, but believe that care should be taken in quantitative analysis.

3. It was further demonstrated here that the alignment of the three orthogonal sections made through the material with respect to the texture is critical to obtain reliable GBPDs. The sections made must be aligned with the principal axes of the texture.

4. We would like to highlight that it is highly desirable to develop a correction algorithm to account for the remaining bias in the GBPD anisotropy level.

Experimental verification of the conclusions described above is also suggested as a possible development of this method. The ground truth might be established using three-dimensional techniques, such as synchrotron X-ray techniques or serial sectioning. EBSD maps from three orthogonal sections should be more straightforward to obtain. The challenge might be to obtain enough boundaries measured for meaningful statistics in stereology.

## Author contributions: CRediT

**Martin Folwarczny:** Investigation, Conceptualisation, Methodology, Writing – original draft, Funding acquisition
**Ao Li:** Investigation, Methodology
**Rushvi Shah:** Investigation, Methodology
**Aaron Chote:** Methodology, Validation
**Alexandra C. Austin:** Methodology
**Yimin Zhu:** Methodology
**Gregory S. Rohrer:** Conceptualisation, Writing – review and editing
**Michael A. Jackson:** Software, Methodology
**Souhardh Kotakadi**: Methodology
**Katharina Marquardt:** Supervision, Conceptualisation, Funding acquisition, Writing – review and editing

## Declaration of competing interest

The authors declare that there are no known competing interests which could influence this work and the contents of this paper.

## Acknowledgements

The authors acknowledge financial support from the EPSRC Centre for Doctoral Training in Nuclear Energy Futures (EP/S023844/1) and the EPSRC Vacation Bursary at Imperial College London for MF and the Royce Undergraduate Internship Scheme in years 2021 for SK and 2022 for RS.

## References

[1] A. Bhattacharya, Y.-F. Shen, C.M. Hefferan, S.F. Li, J. Lind, R.M. Suter, C.E. Krill, G.S. Rohrer, Grain boundary velocity and curvature are not correlated in Ni polycrystals, Science 374 (2021) 189–193. https://doi.org/10.1126/science.abj3210.

[2] P.R. Cantwell, M. Tang, S.J. Dillon, J. Luo, G.S. Rohrer, M.P. Harmer, Grain boundary complexions, Acta Mater. 62 (2014) 1–48. https://doi.org/10.1016/j.actamat.2013.07.037.

[3] G. Dehm, J. Cairney, Implication of grain-boundary structure and chemistry on plasticity and failure, MRS Bull. 47 (2022) 800–807. https://doi.org/10.1557/s43577-022-00378-3.

[4] K. Marquardt, U.H. Faul, The structure and composition of olivine grain boundaries: 40 years of studies, status and current developments, Phys. Chem. Miner. 45 (2018) 139–172. https://doi.org/10.1007/s00269-017-0935-9.

[5] E. Milan, M. Pasta, The role of grain boundaries in solid-state Li-metal batteries, Mater. Futures 2 (2023) 013501. https://doi.org/10.1088/2752-5724/aca703.

[6] A. Sutton, R.W. Balluffi, Mechanical properties of interfaces, in: Interfaces in Crystalline Materials, Corr. ed., Clarendon Press, Oxford, 1996: pp. 704–803.

[7] A. Sutton, R.W. Balluffi, Electronic properties of interfaces, in: Interfaces in Crystalline Materials, Corr. ed., Clarendon Press, Oxford, 1996: pp. 657–703.

[8] H. Davies, V. Randle, Single-section plane assessment in grain boundary engineered brass, J. Microsc. 205 (2002) 253–258. https://doi.org/10.1046/j.1365-2818.2002.00989.x.

[9] S. Toyama, T. Seki, B. Feng, Y. Ikuhara, N. Shibata, Direct observation of space-charge-induced electric fields at oxide grain boundaries, Nat. Commun 15 (2024) 8704. https://doi.org/10.1038/s41467-024-53014-w.

[10] D.A. Molodov, T. Gorkaya, G. Gottstein, Migration of the Σ7 tilt grain boundary in Al under an applied external stress, Scr. Mater. 65 (2011) 990–993. https://doi.org/10.1016/j.scriptamat.2011.08.030.

[11] S.J. Dillon, Y.-F. Shen, G.S. Rohrer, Grain boundary energies in yttria-stabilized zirconia, J. Am. Ceram. Soc. 105 (2022) 2925–2931. https://doi.org/10.1111/jace.18283.

[12] M. Faryna, M. Adamczyk-Habrajska, M. Lubszczyk, Influence of grain boundary plane distribution on ionic conductivity in yttria-stabilized zirconia sintered at elevated temperatures, Archiv.Civ.Mech.Eng 24 (2024) 123. https://doi.org/10.1007/s43452-024-00935-4.

[13] J. Han, V. Vitek, D.J. Srolovitz, Grain-boundary metastability and its statistical properties, Acta Mater. 104 (2016) 259–273. https://doi.org/10.1016/j.actamat.2015.11.035.

[14] G.S. Rohrer, The distribution of grain boundary planes in polycrystals, JOM 59 (2007) 38–42. https://doi.org/10.1007/s11837-007-0114-4.

[15] P. Vonlanthen, B. Grobety, CSL grain boundary distribution in alumina and zirconia ceramics, Ceram. Int. 34 (2008) 1459–1472. https://doi.org/10.1016/j.ceramint.2007.04.006.

[16] V. Randle, ed., Theoretical Aspects of Grain Boundary Geometry I: General Boundaries, in: The Measurement of Grain Boundary Geometry, 1st ed., CRC Press, 1993.

[17] D.M. Saylor, B.S. El-Dasher, B.L. Adams, G.S. Rohrer, Measuring the five-parameter grain-boundary distribution from observations of planar sections, Metall. Mater. Trans. A 35 (2004) 1981–1989. https://doi.org/10.1007/s11661-004-0147-z.

[18] G.S. Rohrer, D.M. Saylor, B.E. Dasher, B.L. Adams, A.D. Rollett, P. Wynblatt, The distribution of internal interfaces in polycrystals, MEKU 95 (2004) 197–214. https://doi.org/10.3139/146.017934.

[19] J.V. Bernier, R.M. Suter, A.D. Rollett, J.D. Almer, High-Energy X-Ray Diffraction Microscopy in Materials Science, Annu. Rev. Mater. Res. 50 (2020) 395–436. https://doi.org/10.1146/annurev-matsci-070616-124125.

[20] H. Pirgazi, K. Glowinski, A. Morawiec, L.A.I. Kestens, Three-dimensional characterization of grain boundaries in pure nickel by serial sectioning *via* mechanical polishing, J. Appl. Crystallogr. 48 (2015) 1672–1678. https://doi.org/10.1107/S1600576715017616.

[21] J.E. Spowart, H.E. Mullens, B.T. Puchala, Collecting and analyzing microstructures in three dimensions: A fully automated approach, JOM 55 (2003) 35–37. https://doi.org/10.1007/s11837-003-0173-0.

[22] M.D. Uchic, Serial Sectioning Methods for Generating 3D Characterization Data of Grain- and Precipitate-Scale Microstructures, in: S. Ghosh, D. Dimiduk (Eds.), Computational Methods for Microstructure-Property Relationships, Springer US, Boston, MA, 2011: pp. 31–52. https://doi.org/10.1007/978-1-4419-0643-4_2.

[23] H.H. Liu, S. Schmidt, H.F. Poulsen, A. Godfrey, Z.Q. Liu, J.A. Sharon, X. Huang, Three-Dimensional Orientation Mapping in the Transmission Electron Microscope, Science 332 (2011) 833–834. https://doi.org/10.1126/science.1202202.

[24] W. Zhu, G. Wu, A. Godfrey, S. Schmidt, Q. He, Z. Feng, T. Huang, L. Zhang, X. Huang, Five-parameter grain boundary character distribution of gold nanoparticles based on three dimensional orientation mapping in the TEM, Scr. Mater. 214 (2022) 114677. https://doi.org/10.1016/j.scriptamat.2022.114677.

[25] M. Groeber, A framework for automated analysis and simulation of 3D polycrystalline microstructures. Part 2: Synthetic structure generation, Acta Mater. 56 (2008) 1274–1287. https://doi.org/10.1016/j.actamat.2007.11.040.

[26] M.A. Groeber, M.A. Jackson, DREAM.3D: A Digital Representation Environment for the Analysis of Microstructure in 3D, Integr. Mater. Manuf. Innov. 3 (2014) 56–72. https://doi.org/10.1186/2193-9772-3-5.

[27] A.T. Motta, A. Couet, R.J. Comstock, Corrosion of Zirconium Alloys Used for Nuclear Fuel Cladding, Annu. Rev. Mater. Res. 45 (2015) 311–343. https://doi.org/10.1146/annurev-matsci-070214-020951.

[28] V.S. Tong, T.B. Britton, Formation of very large 'blocky alpha' grains in Zircaloy-4, Acta Mater. 129 (2017) 510–520. https://doi.org/10.1016/j.actamat.2017.03.002.

[29] Y.N. Wang, J.C. Huang, Texture analysis in hexagonal materials, Mater. Chem. Phys. 81 (2003) 11–26. https://doi.org/10.1016/S0254-0584(03)00168-8.

[30] P. Barberis, Zirconia powders and Zircaloy oxide films: tetragonal phase evolution during 400°C autoclave tests, J. Nucl. Mater. 226 (1995) 34–43. https://doi.org/10.1016/0022-3115(95)00108-5.

[31] P. Bouvier, J. Godlewski, G. Lucazeau, A Raman study of the nanocrystallite size effect on the pressure–temperature phase diagram of zirconia grown by zirconium-based alloys oxidation, J. Nucl. Mater. 300 (2002) 118–126. https://doi.org/10.1016/S0022-3115(01)00756-5.

[32] A. Garner, P. Frankel, J. Partezana, M. Preuss, The effect of substrate texture and oxidation temperature on oxide texture development in zirconium alloys, Journal of Nuclear Materials 484 (2017) 347–356. https://doi.org/10.1016/j.jnucmat.2016.10.037.

[33] F. Bachmann, R. Hielscher, H. Schaeben, Texture Analysis with MTEX – Free and Open Source Software Toolbox, SSP 160 (2010) 63–68. https://doi.org/10.4028/www.scientific.net/SSP.160.63.

[34] J. Ahrens, B. Geveci, C. Law, ParaView: An End-User Tool for Large-Data Visualization, in: Visualization Handbook, Elsevier, 2005: pp. 717–731. https://doi.org/10.1016/B978-012387582-2/50038-1.

[35] T.B. Britton, J. Jiang, Y. Guo, A. Vilalta-Clemente, D. Wallis, L.N. Hansen, A. Winkelmann, A.J. Wilkinson, Tutorial: Crystal orientations and EBSD — Or which way is up?, Mater. Charact. 117 (2016) 113–126. https://doi.org/10.1016/j.matchar.2016.04.008.

[36] G.S. Rohrer, Grain Boundary Crystallography, (2022). http://mimp.materials.cmu.edu/rohrer/gbXstallography/ (accessed August 22, 2023).